\documentclass[conference]{IEEEtran}
\IEEEoverridecommandlockouts
\usepackage{cite}
\usepackage{marvosym}
\usepackage{amsmath,amssymb,amsfonts}
\usepackage{graphicx}
\usepackage[noend]{algorithmic}
\usepackage{url}
\usepackage{algorithm}
\usepackage{algorithmic}

\usepackage{subfigure}
\usepackage{textcomp}
\usepackage{xcolor}

\def\BibTeX{{\rm B\kern-.05em{\sc i\kern-.025em b}\kern-.08em
    T\kern-.1667em\lower.7ex\hbox{E}\kern-.125emX}}
\begin{document}

\title{Computation Resource Leasing for Priority Aggregation Local Computing Network}
\author{\IEEEauthorblockN{1\textsuperscript{st} Chao Ren}
\IEEEauthorblockA{\textit{University of Science} \\
\textit{and Technology Beijing} \\
Beijing, China \\
chaoren@ustb.edu.cn}

\and

\IEEEauthorblockN{2\textsuperscript{st} Jingze Hou}
\IEEEauthorblockA{\textit{University of Science} \\
\textit{and Technology Beijing} \\
Beijing, China \\
jingzehou@163.com}

\and

\IEEEauthorblockN{3\textsuperscript{rd} Haijun Zhang\textsuperscript{\Letter}}
\IEEEauthorblockA{\textit{University of Science} \\
\textit{and Technology Beijing}\\
Beijing, China \\
haijunzhang@ieee.org}
\and

\IEEEauthorblockN{4\textsuperscript{th} other coauthors}
\IEEEauthorblockA{\textit{other institutions} \\
\textit{}\\
 \\
}

}

\maketitle

\begin{abstract}
In large scale smart edge networks, computation resource is generally underutilized due to the uneven distribution of computation resource  in time and space domain. This may correspond to a simple fact that no device is capable for `storing' and `exchanging' idle computation resource. Thus, this paper proposes a computation resource leasing (CRL) concept  using priority as an intermediary to restore and exchange the permission for computation resource for priority aggregation local computing network (PALCN). Each device in PALNC is able to gain priority as a reward for leasing its computing resource to others.
CRL also offers a priority oriented algorithm to match the computation request with idle source nodes and a priority management model. Our analysis and numerical results show that the system can efficiently utilize local idle computation sources over time and space domain and filtrate the big task that local computation can not finish.
\end{abstract}


\section{Introduction}
 The explosive growth is observed both in the number of mobile devices and in the amount of related computation. By the year 2020, total quantity of world-wide mobile devices would be 75 billion\cite{white} with  a traffic volume exceeding 24.3 exabytes/month, which also leads to high computation load of the same order. However, some devices are idle considering its utilization in time and space domain, e.g., smart devices plugging in charging ports, private edge IoT servers at night, and idle servers some distance away.
 Although there may be no urging tasks requiring computation resource, these idle devices may encounter heavy computing load in the future.
 Can we `borrow' the idle computation resource from the past to help release some incoming heavy pressure?
  
 Unfortunately, computation resource cannot be stored or exchanged, due to its real-time computing characteristic. Thus, we could define the aforementioned problem as uneven distribution of computation resource in time and space domain with real-time computing nature.
 To solve this problem, PALCN is defined to group some devices having low latency to other devices, and they are willing to share and request computation resources with high priority devices.
In a PALCN, there may be idles computation resources, real-time computation tasks, future tasks, and devices sharing their computation resource for potential rewards.
The uneven distribution of computation resource in time and space may be materialize into unfairness in giving and taking for devices in PALNC.
Therefore, the key problem is to find a  mechanism to fully utilize  local resource with a new CRL concept matching real-time/future tasks with idle or semi-idle\footnote{Semi-idle means the devices that are not in full load computing e.g. computers with CPU utilization of only 30 percent } devices.

To release the pressure of local computation tasks, many concepts have been applied in reality. The authors in\cite{6787113} proposed an algorithm to offload the local tasks to mobile cloud based on game theory. Reference\cite{7845499} introduced an algorithm that offload the local tasks to edge computing layer to decrease the computing energy consumption. The authors of reference\cite{6983145} propose an algorithm to offload the tasks based on increasing utilization of the network.  
These related works are able to efficiently utilize the released computation resource from other layers to balance the uneven distribution of computation resources in space, which are also known as\textit{ task offloading}\cite{8314696,8515736,8359090} or \textit{task migration}\cite{8648762,7064025,7518364}. 
However, in local network, task offloading or migration may not solve the resource constrains in time domain.
Considering a group of local devices may cooperate and accumulate their `kindness', they may have high priority to offload their future tasks to other idle or semi-idle devices cross-time.
Could these devices work in a satisfying mechanism managing all the sharing, priority exchanging and restoring functions?

Thus, a new CRL concept is proposed in this paper, where the CRL member devices lease computing resource to other PALCN devices and  obtain priority for their own computing.
Using priority, the computation priority can be thus be accumulated, which facilitates the task offloading regardless time and space constrains, i.e., the  idle computation resource restoring and exchanging problem is converted to priority accumulation and consumption problem.
To match the computation tasks with idle computation within PALCN, algorithms are proposed and sort big task with matching methods. Moreover, a settlement system is proposed to manage the priority in leasing.
Compared with task offloading to cloud or edge layer, CRL is able to fully utilize the local computation resource and aggregate the priority for future heavy tasks.

In summary, The distinctive feathers and contributions of this paper are as follows:
\begin{itemize}
\item  This paper proposes a new concept named CRL to convert the across-time invocation of the computation. 
\item This paper proposes PALCN to redivide the devices in smart edge networks and is suitable for the CRL model. Finally, the conclusion is in section V.   
\end{itemize}

The remainder of this paper is organized as follows. In section II. The proposed matching algorithm is described in detail in section III. Performance analyses are presented in section IV.

\begin{table*}[!h]
	\caption{Important notations.}\label{table_notation}
	\centering
	\begin{tabular}{ | p{1.7cm} | p{14.0cm}|}
		\hline
		Notation & Explanation \\ \hline
     $D_{Ti}$
      & Deadline of the $i^{th}$ task, the unit of it is second. It describe that the computing source must finish $i^{th}$ task before $D_{Ti}$ 	\\ \hline 
		$A_{Ti}$
		& It denotes the total
number of CPU cycles required to accomplish the computation task.\\ \hline
	$V_{Ti}$	
& A constant that describe the volume of $i^{th}$ task.\\ \hline
$P_{Ti}$
& A constant to describe the value of $i^{th}$ task task per bit.\\ \hline
	$N_{Ti}$
&	The number of $i^{th}$ task. \\  \hline
 $E_{Tj}$
 & The time that $j^{th}$ source node can compute for the task, the unit of it is second.\\ \hline
 $ Cal_{Tj}$
 & The value to describe the $j^{th}$ source node’s computing ability, the unit of it is bit per second  \\ \hline
    $M_{j}$
& The number of the $j^{th}$ source node. \\ \hline
$Prefer_(j,i)$
&  The time that $i^{th}$ task can be finished by $j^{th}$ source node. \\ \hline
$B_{j}$
& The amount of the priority that exchange in the process of CRL. \\ \hline
$R$
& A constant that describe the conversion rate between task value and token.\\
\hline
$W$
& The times that CRL algorithm repeat.\\
\hline 
$\gamma^{T}$
& The weight of of value of the task per bit.\\
\hline
$\gamma^{p}$
& The weight of priority.\\
\hline
	\end{tabular}
\end{table*}

\section{SYSTEM MODEL}
\begin{figure}[htbp]
\center
\centering\includegraphics[scale=0.4]{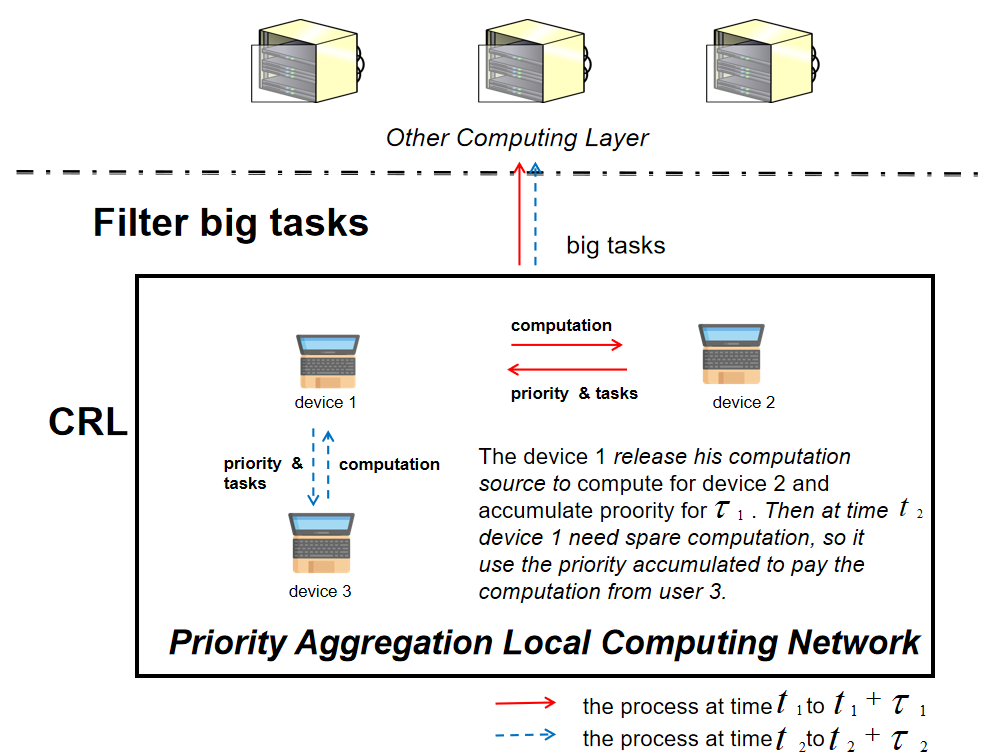}
\caption{Simplified system model : consisted of the process CRL and process filter big tasks.}
\label{fig_systemmodel}
\end{figure}

In this model, we define some devices that are logically adjacent or have time delay lower than $\tau$ with each other as members in a  PALCN. There are some idle or semi-idle devices (provider) and some devices  that need extra computation (receiver) in a PALCN (Fig. 1). The process that the provider compute for receiver is also the process in which the receiver pays the priority to the provider and the provider accumulates its priority. We call above process as CRL. When device 1 realizes the transformation from the provider at time $t_1$ to the receiver at time $t_2$, we think that it realizes the call of computation across-time. In addition, the system should have ability to filter the big tasks that local computation source can not be finished. Thus, the system model is proposed as follows (see Fig. \ref{fig_systemmodel}).

We regard the CRL as process consist of matching process and priority settlement process. So we model the matching process and settlement  respectively.

\subsection{Model of Matching Computation Source}\label{AA}
Suppose there are $n$ tasks need to be computed and $m$ idle or semi-idle devices in the PALCN at one time.
Regard $m$ idle devices as $m$ computation sources with different computation ability. For every task, there are some parameters. Regard each task as a column vector, these parameters as the elements of the column vector. So a task can be expressed as ($ 1 \leq $i$ \leq n $,$~$i$\in\ N$):

\begin{align}
	\textbf{\text{T}}_{i}&=\left [
	\begin{matrix}
		D_{Ti}\\
		A_{Ti}\\
	    V_{Ti}\\
		P_{Ti}\\
		N_{Ti}\\
		\textit{priority}_{i}\\
	\end{matrix}\right ], \label{equ_h}
\end{align}
where the $D_{Ti}$ means the deadline of task $i$. $A_{Ti}$ denotes the total
number of CPU cycles required to accomplish the computation task. $V_{Ti}$ denotes the value of tasks $i$. $P_{Ti}$ denotes the matching priority of task $i$. $N_{Ti}$ denotes the number of task $i$. $\textit{priority}_{i}$ denotes the priority that the owner of the task $i$ have.

Each tasks have different matching priority. We need to match the task in high matching priority to computing source previously. In this model, matching priority is expressed as:

\begin{equation}
	P_{Ti}=\gamma^{T}\frac{V_{Ti}}{A_{Ti}}+\gamma^{P}priority_{i} , 
	\label{equ:antenna_rule}
\end{equation}
where $ 0 \leq \gamma^{T},\gamma^{P} \leq 1 $ denotes the weight of value of the task per bit and its priority. The task has a higher priority means that the device of the task has completed more computation for others before.

For every computation source, there are some parameters. Regard each source as a column vector, these parameters as the elements of the column vector. So a source can be expressed as($ 1 \leq $j$ \leq m $,$~$j$\in\ N$):

\begin{align}
	\textbf{\text{S}}_{j}&=\left [
	\begin{matrix}
		E_{Tj}\\
		Cal_{Tj}\\
	    M_{j}\\
	\end{matrix}\right ], \label{equ_h}
\end{align}
where $E_{Tj}$ denotes the time that $S_{j}$ be idle, the unit of it is second.
$Cal_{Tj}$ denotes the computing ability of the source $j$, and the unit of it is the number of CPU cycle it can finish per second. $M_{j}$ denotes the number of that source.

Make $n$ tasks to generate a 6$\times$$n$ \textit{all tasks matrix} :

\begin{align}
	\textbf{\text{T}}_{all}&=\left [
	\begin{matrix}
		D_{T1}&\cdots&\cdots& D_{Tn}\\
		A_{T1}&\cdots&\cdots& A_{Tn}\\
        V_{T1}&\cdots&\cdots& V_{Tn}\\
        P_{T1}&\cdots&\cdots& P_{Tn}\\
        N_{T1}&\cdots&\cdots& N_{Tn}\\
        priority_{1}&\cdots&\cdots&priority_{n}\\
	\end{matrix}\right ]. \label{equ_h}
\end{align}

Make $m$ sources to generate a 3$\times$$m$ \textit{all source matrix}:

\begin{align}
	\textbf{\text{S}}_{all}&=\left [
	\begin{matrix}
		E_{T1}&\cdots&\cdots& E_{Tm}\\
		Cal_{T1}&\cdots&\cdots& Cal_{Tm}\\
        M_{T1}&\cdots&\cdots& M_{Tm}\\
	\end{matrix}\right ]. \label{equ_h}
\end{align}

This paper introduces \textit{priority} into the model to generate the \textit{priority matrix}. Rearrange the column vector of \textit{all task matrix} based on $P_{Ti}$ in descending order. Using bubble sort to exchange column task vectors' order to make tasks with higher \textit{priority}  more previous. Reorder the the subscript of all tasks and generate a new matrix named \textit{priority matrix}:

\begin{align}
	\textbf{\text{T}}_{Pall}&=\left [
	\begin{matrix}
		D_{T1}&\cdots&\cdots& D_{Tn}\\
		A_{T1}&\cdots&\cdots& A_{Tn}\\
        V_{T1}&\cdots&\cdots& V_{Tn}\\
        P_{T1}&\cdots&\cdots& P_{Tn}\\
        N_{T1}&\cdots&\cdots& N_{Tn}\\
        priority_{1}&\cdots&\cdots&priority_{n}\\
	\end{matrix}\right ]. \label{equ_h}
\end{align}

The CRL process translates into matching the column vectors of \textit{priority matrix} and the column vectors of \textit{all source matrix}. We can regard above process as matching the appropriate provider for receiver.

The another process of system is priority settlement, it can be modeled as follow:
\subsection{Model of Priority Accumulation and Payment}
Each time a CRL is completed, the priority transaction process is settled, and the receiver pays priority to the provider. We define a constant $R$ to describe the conversion rate between priority and task volume.

\begin{equation}
	B_{i}=(\gamma^{N}\times V_{Ti}+\gamma^{M}\times priority_{i})\times R , 
	\label{equ:antenna_rule}
\end{equation}
where $ 0 \leq \gamma^{N},\gamma^{M} \leq 1 $ denotes the weight of volume of the task and the amount of priority it own. $B_{i}$ denotes amount of priority that receiver should pay to the provider. Each time a CRL process is completed, receiver exchange $B_{i}$ priority with provider.

\section{CRL ALGORITHM AND PRIORITY SETTLEMENT ALGORITHM }

The model has transform the CRL to a linear one, which make it easy to match the most preferred computation source for each task and exchange priority. Task in higher $P_{i}$ select the resources with the more preferred to exchange priority. In this way, a positive cycle is formed. The more a device computes for other devices in a PALCN, the higher priority it have, and when it needs additional computation, it can choose resources for priority exchange previously. Tasks without resources are placed in the task matrix at the next moment to re-match resources, and the matching process is repeated $W$ times. A task still with no resources after $W$ matches is treated as big task, and upload it to other computing layer e.g. cloud.

\subsection{CRL Algorithm}

\begin{algorithm}[tbh!] 
	\caption{Generate the \textit{prefer matrix}} 
	\begin{algorithmic}[1]
   \REQUIRE$\textbf{\text{S}}_{all}=\left [
      \begin{matrix}
		E_{T1}&\cdots&\cdots& E_{Tm}\\
    	Cal_{T1}&\cdots&\cdots& Cal_{Tm}\\
      M_{T1}&\cdots&\cdots& M_{Tm}\\
   	\end{matrix}\right ] \label{equ_h}	$

	$\textbf{\text{T}}_{Pall}=\left [
	\begin{matrix}
		D_{T1}&\cdots&\cdots& D_{Tn}\\
		A_{T1}&\cdots&\cdots& A_{Tn}\\
        V_{T1}&\cdots&\cdots &V_{Tn}\\
        P_{T1}&\cdots&\cdots &P_{Tn}\\
        N_{T1}&\cdots&\cdots &N_{Tn}\\
         priority_{1}&\cdots&\cdots&priority_{n}\\
	\end{matrix}\right ] \label{equ_h}$
\ENSURE 

   $prefer_{(j,i)}=\frac{Cal_{Tj}}{A_{Ti}}$

      $\textbf{\text{Pref}}=\left [
      \begin{matrix}
		prefer_{(1,1)} &\cdots&\cdots &prefer_{(1,n)}\\
    	prefer_{(2,1)} &\cdots&\cdots &prefer_{(2,n)}\\
        \vdots &\ddots&\ddots&\vdots\\
        prefer_{(m,1)} &\cdots&\cdots &prefer_{(m,n)}\\
   	
   	\end{matrix}\right ] \label{equ_h}	$

	\end{algorithmic} 
\end{algorithm}

After that the \textit{priority matrix} shows a descending priority task order from left to right.
For the task $T_{i}$, if a idle source $S_{J}$ can finish the task on time, the $S_{j}$ need to have enough computation and can finish before the $D_{Ti}$ of $T_{i}$, so it must meet($1 \leq i \leq n$, $1 \leq j \leq m$):

\begin{equation}
   A_{Ti}\leq Cal_{Tj}\times E_{Tj},
	\label{equ:antenna_rule}
\end{equation}

\begin{equation}
  \frac{A_{Ti}}{Cal_{Tj}}\leq D_{Ti}.
	\label{equ:antenna_rule}
\end{equation}

\begin{itemize}
\item  Case I: $S_{j}$ can finish $T_{i}$ :

  \begin{equation}
   prefer_{(j,i)}=\frac{Cal_{Tj}}{A_{Ti}}.
	\label{equ:antenna_rule}
   \end{equation}
\item  Case II: $S_{j}$ can not finish $T_{i}$ :
    \begin{equation}
   prefer_{(j,i)}=0.
	\label{equ:antenna_rule}
   \end{equation}
\end{itemize}

\begin{algorithm}[tbh!] 
	\caption{Generate the \textit{Match matrix}} 
	\begin{algorithmic}[1]
   \REQUIRE$\textbf{\text{Pref}}=\left [
      \begin{matrix}
		prefer_{(1,1)} &\cdots&\cdots &prefer_{(1,n)}\\
    	prefer_{(2,1)} &\cdots&\cdots &prefer_{(2,n)}\\
        \vdots &\ddots&\ddots&\vdots\\
        prefer_{(m,1)} &\cdots&\cdots &prefer_{(m,n)}\\
   	
   	\end{matrix}\right ] \label{equ_h}	$

\ENSURE 

\STATE Recording the highest element in the first vector column, and set other elements in row vector as zero which means that source has been selected by the task.
\STATE Recording the highest element in the second vector column, and set other elements in row vector as zero which means that source has been selected by the task.
\STATE Repeat the process above.
 
	
\STATE	$\textbf{\text{Match}}=\left [
      \begin{matrix}
		N_{(t1)} &\cdots&\cdots &N_{(Tn)}\\
    	M_{(j)} &\cdots&\cdots &M_{(k)}\\
   	\end{matrix}\right ] \label{equ_h}	$
	
	\end{algorithmic} 
\end{algorithm}

For each provider, CRL can match it with a receiver to help it storage its idle computation as priority. At some time in the future, when the device needs extra computation, it can exchange the stored priority to computation from idle devices. Thus CRL realize balancing the uneven distribution of computation resources across-time. The vector column of \textit{Match matrix} show the priority exchange relation between providers and receivers. 

In actual use, there may be some tasks are too big to finished by local computing source, so filter them and upload to other computing layer is necessary.

\begin{algorithm}[tbh!] 
	\caption{Filter big tasks} 
	\begin{algorithmic}[1]
   \REQUIRE$\textbf{\text{Pref}}=\left [
      \begin{matrix}
		prefer_{(1,1)} &\cdots&\cdots &prefer_{(1,n)}\\
    	prefer_{(2,1)} &\cdots&\cdots &prefer_{(2,n)}\\
        \vdots &\ddots&\ddots&\vdots\\
        prefer_{(m,1)} &\cdots&\cdots &prefer_{(m,n)}\\
   	
   	\end{matrix}\right ] \label{equ_h}	$

\ENSURE

\STATE Recording the highest element in the first vector column, and set other elements in row vector as zero which means that source has been selected by the task.
\STATE Recording the highest element in the second vector column, and set other elements in row vector as zero which means that source has been selected by the task.
\STATE If all the elements in a vector column of the matrix, which means the there is no computing source can finish the task that the vector column represented. So, put that task into \textit{task list } at next time.
\STATE If there is still no source can finish it after repeating $W$ times, regard it as big task and upload it to cloud to finish.
	\end{algorithmic} 
\end{algorithm}

If all the computing resources the task can choose cannot finish it, there are two possible remote causes

\begin{itemize}
\item  Case I: Its priority is too low to choose the computing resources previously.

\item  Case II: There is no computing source have enough computation to finish it.

For whatever reason, we regard it as a big task to avoid $\vartheta{(m \times n) }$ complexity of the algorithm.

\end{itemize}

\subsection{Priority Settlement  }

 After each process of exchanging, provider and receiver both should 
update their priority.

\begin{algorithm}[tbh!] 
	\caption{Priority settlement} 
	\begin{algorithmic}[1]
   \REQUIRE$\textbf{\text{Match}}=\left [
      \begin{matrix}
		N_{(t1)} &\cdots&\cdots &N_{(Tn)}\\
    	M_{(j)} &\cdots&\cdots &M_{(k)}\\
   	\end{matrix}\right ] \label{equ_h}	$

$	B_{i}=(\gamma^{N}\times V_{Ti}+\gamma^{M}\times priority_{i})\times R $ 	

\ENSURE from i=1 to i=n 

\textit{$priority_{i}$}=\textit{$priority_{i}$}$-$$B_{i}$

from j=1 to j=m

\textit{$priority_{j}$}=\textit{$priority_{j}$}$+$$B_{j}$

	\end{algorithmic} 
\end{algorithm}

Priority settlement can help to exchange the priority between provider and receiver by recording the amount of the priority transaction and the priority their hold after the transaction.

\section{Numerical Results}

In the following simulations, we will compare CRL with cloud offloading\cite{6787113} in terms of \textit{cross-time distribution of idle computing resources} , and migration data of computing tasks to measure the performance of CRL. The simulation intercepts a short period of time in CRL process. The main notations of the simulation are as follow: 

$\gamma^{T}$=0.5 ,
$\gamma^{P}$=0.5

\subsection{Cross-Time Distribution of Idle Computation Resources}
Cross-time distribution means the amount of idle computing resources changes with time. The less idle computing resources, the higher the utilization of resources in this network.

\begin{figure}[htbp]
\center
\centering\includegraphics[scale=0.5]{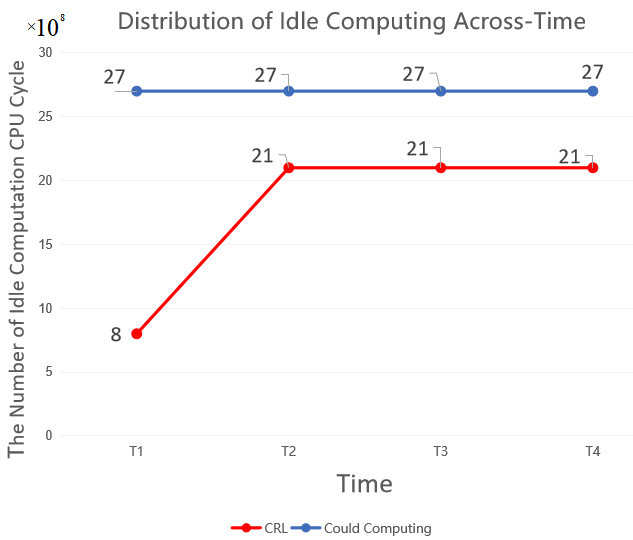}
\caption{Cross-Time Distribution of Idle Computation Resources
}
\label{fig}
\end{figure}

It can be seen from the simulation that CRL can reduce \textit{Cross-Time Distribution of Idle Computation Resources} to improve the utilization of devices in the PALCN. Because simulation is a short period of time in the process of CRL interception, the \textit{across-time distribution of Idle Computation Resources} of cloud computing changes to a constant with time. Compared with cloud computing, CRL can make full use of the resources of PALCN to compute the internal tasks in PALCN.
 
\subsection{Data of Computing Tasks  Migration}

Cloud computing will transfer all local tasks to the cloud server, which we call task migration. However, CRL will only transfer big tasks that cannot be completed locally to the cloud. We can measure the filtering effect of CRL on large tasks in PALCN by monitoring the amount of migrated tasks.

\begin{figure}[htbp]
\center
\centering
\includegraphics[scale=0.5]{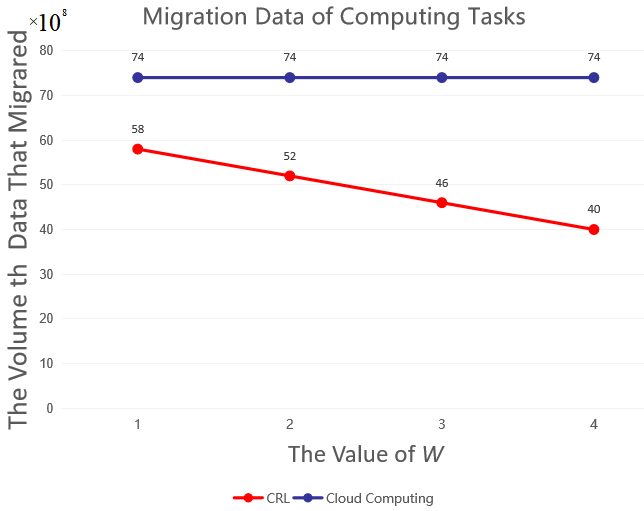}
\caption{Migration Data of Computing Tasks
}
\label{fig}
\end{figure}

Numerical result shows that comparing with cloud computing, CRL can reduce the amount of task migration, and with the $W$ increasing, the migration task reduced. This also shows that when the number of CRL process repetitions increases, the filtering effect for big tasks is better.

\section{Conclusion}

In this paper, we consider the problem that unbalanced distribution of computation source across-time. We propose new concepts named PALCN and CRL to solve that problem. CRL can reduce the pressure of cloud by finishing some tasks in PALCN using local idle computation source. The most important contribution of CRL is that it realize the call of computation across-time by exchanging priority, which means realizing the storage of computation.

For the future work we are considering more general cases that expand the scope of the network, it is not limited to the PALCN, so that the devices with higher delay can also call computation resources across-time through priority exchange.

\bibliographystyle{IEEEtran}
\bibliography{main}
\vspace{12pt}

\end{document}